\documentclass[%
 reprint,
superscriptaddress,
 amsmath,amssymb,
 aps,
pra,
floatfix,
breaklinks=true,
]{revtex4-2}

\usepackage{graphicx}
\usepackage{dcolumn}
\usepackage{bm}
\usepackage{times}
\usepackage{xcolor}
\usepackage{orcidlink}
\newcommand{\ket}[1]{\lvert#1\rangle} 

\begin{document}
\title{Adaptive Interpolating Quantum Transform: A Quantum-Native Framework for Efficient Transform Learning}

\author{Gekko Budiutama\orcidlink{0000-0001-8904-9983}}
\email{bgekko@quemix.com}
\affiliation{Quemix Inc.,
Taiyo Life Nihonbashi Building,
2-11-2,
Nihonbashi Chuo-ku, 
Tokyo 103-0027,
Japan}
\affiliation{Department of Physics, 
The University of Tokyo, 
Tokyo 113-0033, 
Japan}

\author{Shunsuke Daimon\orcidlink{0000-0001-6942-4571}}
\email{daimon.shunsuke@qst.go.jp}
\affiliation{Quantum Materials and Applications Research Center, National Institutes for Quantum Science and Technology, Tokyo, 152-8550, Japan}

\author{Hirofumi Nishi\orcidlink{0000-0001-5155-6605}}
\affiliation{Quemix Inc.,
Taiyo Life Nihonbashi Building,
2-11-2,
Nihonbashi Chuo-ku, 
Tokyo 103-0027,
Japan}
\affiliation{Department of Physics, 
The University of Tokyo, 
Tokyo 113-0033, 
Japan}

\author{Ryui Kaneko\orcidlink{0000-0001-7994-6381}}
\affiliation{Physics Division, Sophia University, Chiyoda, Tokyo 102-8554, Japan}

\author{Tomi Ohtsuki\orcidlink{0000-0002-4069-6917}}
\affiliation{Physics Division, Sophia University, Chiyoda, Tokyo 102-8554, Japan}

\author{Yu-ichiro Matsushita\orcidlink{0000-0002-9254-5918}}%
\affiliation{Quemix Inc.,
Taiyo Life Nihonbashi Building,
2-11-2,
Nihonbashi Chuo-ku, 
Tokyo 103-0027,
Japan}%
\affiliation{Department of Physics, 
The University of Tokyo, 
Tokyo 113-0033, 
Japan}
\affiliation{Quantum Materials and Applications Research Center, National Institutes for Quantum Science and Technology, Tokyo, 152-8550, Japan}%

\begin{abstract}
Machine learning on quantum computers has attracted attention for its potential to deliver computational speedups in different tasks. However, deep variational quantum circuits require a large number of trainable parameters that grows with both qubit count and circuit depth, often rendering training infeasible. In this study, we introduce the Adaptive Interpolating Quantum Transform (AIQT), a quantum-native framework for flexible and efficient learning. AIQT defines a trainable unitary that interpolates between quantum transforms, such as the Hadamard and quantum Fourier transforms. This approach enables expressive quantum state manipulation while controlling parameter overhead. It also allows AIQT to inherit any quantum advantages present in its constituent transforms. Our results show that AIQT achieves high performance with minimal parameter count, offering a scalable and interpretable alternative to deep variational circuits.
\end{abstract}

\maketitle
\section{Introduction}
Deep neural networks have demonstrated exceptional performance across computer vision, natural language processing, and many other domains, but their success often comes at the cost of very high computational resources, including large model depths and significant parameter counts \cite{resnet50, vaswani2023, brown2020, kaplan2020}. To address this, there has been sustained interest in integrating well-defined mathematical transforms that can reduce network complexity while introducing meaningful inductive biases \citep{https://doi.org/10.1049/sil2.12109, Yi2023ASO}. Classic examples include the discrete Fourier transform for spectral analysis \citep{guibas2022, tancik2020fourier, lou2021fnetarmixingtokensautoregressive, wang2024timemixer}, the discrete cosine transform for image processing \citep{Xu2020, 7415375, Karaoglu2023, Lee2024, SU2024106139}, wavelet transforms for multi-resolution representations \citep{liu2019multi, WaveletFlow}, and Legendre transforms for solving differential equations and scientific computing tasks \citep{LDNN2021, Khan2022, Yang2019}. These transforms are routinely used in deep learning pipelines to compress, decorrelate, or efficiently mix features, and have even been used to replace self-attention in transformers with spectral-domain mixing for natural language processing \citep{leethorp2022fnetmixingtokensfourier, Scribano2023}. Furthermore, our recent work has proposed an interpolating framework, which learns data-driven combinations of multiple transforms to provide adaptable, efficient representations with minimal additional parameters \cite{budiutama2025}.

Quantum machine learning aims to harness quantum computational resources to accelerate tasks such as classification, feature extraction, and generative modeling \cite{biamonte2017, schuld2018}. Inspired by their classical counterparts, various forms of deep quantum neural networks have been proposed, aiming to leverage the expressive power of parameterized quantum circuits \cite{Pan2023, Beer2020, GarcaMartn2025}. However, these architectures face several unique challenges. Quantum resources, such as the number of gates and circuit depth, are inherently expensive and constrained by hardware limitations \cite{Preskill2018}. Optimizing large numbers of parameters in quantum circuits is also resource intensive, requiring repeated quantum measurements that incur significant sampling costs \cite{Kandala2017}. Moreover, training deep quantum neural networks with gradient-based methods is particularly difficult due to issues like barren plateaus, where gradients vanish exponentially with system size, making learning infeasible \cite{mcclean2018, cerezo2021}. Several approaches have been proposed to mitigate these challenges by improving parameter efficiency through better initialization \cite{Grant2019initialization, sugawara2025embeddingtreetensornetworks}, learning strategies \cite{Skolik2021} and post-processing protocols \cite{budiutama2024}. However, it remains critical to develop quantum-native models that achieve efficient circuit designs with minimal parameter counts, while preserving expressive capacity and reducing resource requirements.

In this study, we introduce the Adaptive Interpolating Quantum Transform (AIQT), a quantum-native framework for efficient quantum representation learning. Expanding on our previous work \cite{budiutama2025} to quantum domain, AIQT constructs a trainable unitary that interpolates between known quantum transforms. By learning within a structured subspace of unitaries, AIQT provides meaningful inductive bias and requires only a small number of parameters. Importantly, AIQT inherits the quantum advantage of its constituent unitaries while enabling continuous, learnable interpolation between them. 

\begin{figure}[ht]
    \centering
    \includegraphics[width=0.45\textwidth]{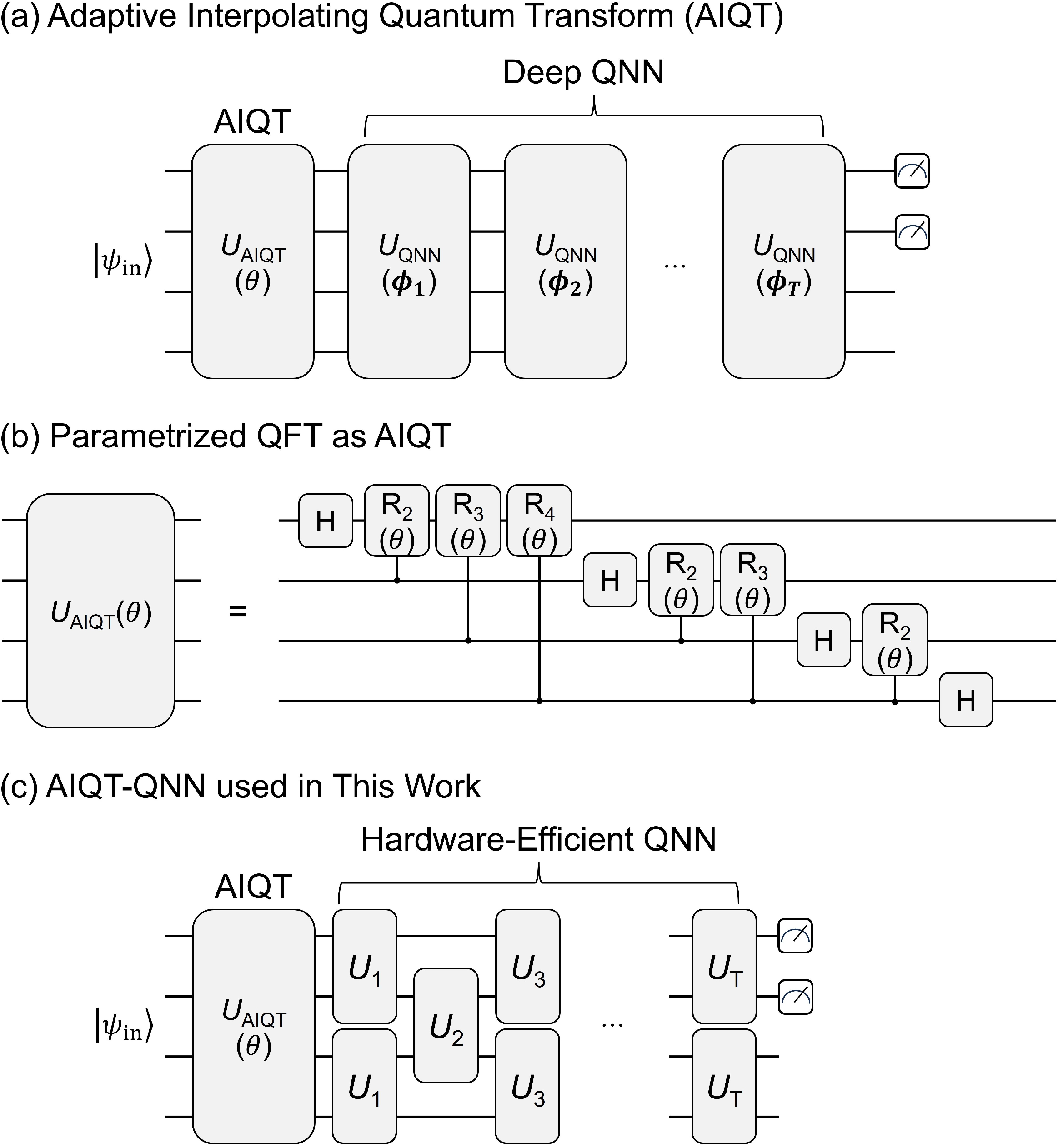}
    \caption{(a) A schematic illustration of the Adaptive Interpolating Quantum Transform (AIQT) as a low-cost adaptive unitary operator. Here, $\phi_1, \dots, \phi_{T}$ denote the trainable parameters of the QNN.
    (b) Parametrized QFT as an instance of AIQT. Here, a trainable global parameter $\theta$ is introduced into the QFT phase structure, enabling smooth interpolation between Hadamard ($\theta = 0$) and full QFT ($\theta = 2\pi$). The controlled-rotation gate, denoted $CR_k(\theta)$, applies a phase of $\theta/2^k$ ($CR_k(\theta) = \mathrm{diag}(1, 1, 1, e^{i \theta / 2^k})$). (c) Architecture of the AIQT-QNN used in this work, combining AIQT-based input transformation with a hardware-efficient QNN ansatz.}
    \label{fig1}
\end{figure}

\section{The Adaptive Interpolating Quantum Transform}
The Adaptive Interpolating Quantum Transform (AIQT) is defined as a trainable unitary operator \(U_{\mathrm{AIQT}}(\boldsymbol{\theta})\) that smoothly interpolates between well-defined unitary transforms \(U_A, U_B, \ldots\). This interpolation is governed by a small set of global or structured trainable parameters \(\boldsymbol{\theta}\). The fundamental principle of the AIQT is to provide a meaningful and structured inductive bias for a quantum model, allowing it to learn the optimal transformation for a given task from a constrained subspace of unitary operations.

The AIQT recovers its constituent transforms in specific regions of the parameter space. For example, when interpolating between two unitaries $U_A$ and $U_B$:
\begin{gather}
    \lim_{\boldsymbol{\theta} \to \boldsymbol{\theta}_A} U_{\mathrm{AIQT}}(\boldsymbol{\theta}) = U_A, \\
    \lim_{\boldsymbol{\theta} \to \boldsymbol{\theta}_B} U_{\mathrm{AIQT}}(\boldsymbol{\theta}) = U_B.
\end{gather}
Crucially, \(U_{\mathrm{AIQT}}(\boldsymbol{\theta})\) remains unitary for all values of \(\boldsymbol{\theta}\), ensuring that the transformation is physically valid. This construction allows the model to inherit the known quantum advantages of its constituent transforms (e.g., the speedup associated with \(U_B\)) while retaining the flexibility to simplify itself towards \(U_A\) if the data or problem structure demands it. 

While the above describes interpolation between two unitaries, the AIQT framework can be naturally extended to interpolate among multiple target transforms by appropriately designing the parameter set \(\boldsymbol{\theta}\). Importantly, AIQT achieves adaptivity through a strategically minimal set of trainable parameters embedded within the unitary structure (Fig.~\ref{fig1}(a)).

In this work, we present a concrete instantiation of the AIQT that interpolates between the Hadamard transform ($U_{\mathrm{H}}$) and the Quantum Fourier Transform ($U_{\mathrm{QFT}}$), as shown in Fig.~\ref{fig1}(b). Here, we define $U_A \equiv U_{\mathrm{H}}$ and $U_B \equiv U_{\mathrm{QFT}}$. The standard circuit for the QFT consists of initial Hadamard gates on each qubit, followed by a series of controlled phase-rotation gates, $CR_k$, that apply a phase of $2\pi/2^k$ ($CR_k = \mathrm{diag}(1, 1, 1, e^{i 2\pi / 2^k})$). To construct our AIQT, we parameterize these phase rotations with a single global parameter $\theta$. The controlled-rotation gate, now denoted $CR_k(\theta)$, applies a phase of $\theta/2^k$ ($CR_k(\theta) = \mathrm{diag}(1, 1, 1, e^{i \theta / 2^k})$). This parameterization achieves the desired interpolation:
\begin{itemize}
    \item When $\theta = 0$, all controlled-rotation gates apply a phase of 0, becoming identity gates. The circuit thus simplifies to only the Hadamard gates, recovering the Hadamard transform: $U_{\mathrm{AIQT}}(0) = U_{\mathrm{H}}$.
    \item When $\theta = 2\pi$, the phases become $2\pi/2^k$, recovering the exact circuit for the standard Quantum Fourier Transform: $U_{\mathrm{AIQT}}(2\pi) = U_{\mathrm{QFT}}$.
\end{itemize}

The AIQT is employed as an adaptive global mixing layer within a larger variational quantum model, as depicted in Fig.~\ref{fig1}(a). An input state $|\psi_{\mathrm{in}}\rangle$ first passes through the AIQT, $U_{\mathrm{AIQT}}(\theta)$, and the resulting state is then processed by a shallow, task-specific Quantum Neural Network (QNN), denoted $U_{\mathrm{QNN}}(\boldsymbol{\phi})$. The final output state of the model just before measurement is given by:
\begin{gather}
    \label{eq:in_out}
    |\psi_{\mathrm{out}}\rangle = U_{\mathrm{QNN}}(\boldsymbol{\phi}) U_{\mathrm{AIQT}}(\theta) |\psi_{\mathrm{in}}\rangle.
\end{gather}
Here, $\boldsymbol{\phi}$ represents the vector of trainable parameters within the QNN. This architecture enables the AIQT to act as a learnable, task-adaptive preprocessing layer that enhances the expressive power of the shallow QNN while maintaining a minimal parameter count.

\section{Methods}
To demonstrate the effectiveness of AIQT, we evaluate AIQT-QNN models on an open-boundary system governed by a Hamiltonian exhibiting $\mathbb{Z}_2 \times \mathbb{Z}_2^{T}$ symmetry:
\begin{equation}
    \label{eq:energy_2}
    H = g_{zxz} \sum_{i=2}^{N-1} Z_{i-1} X_i Z_{i+1} - g_x \sum_{i=1}^{N} X_i - g_{zz} \sum_{i=1}^{N-1} Z_i Z_{i+1},
\end{equation}
where $g_{zxz}$, $g_x$, and $g_{zz}$ are tunable Hamiltonian parameters that satisfy $g_{zxz} + g_x + g_{zz} = 4$ \cite{smith_crossing_2022}. This model was chosen for benchmarking due to its exact solvability. Depending on the parameter values, its ground state can realize one of three distinct phases: a symmetry-protected topological (SPT) phase characterized by a cluster state, a trivial phase, or a symmetry-broken (SB) phase. To construct the dataset for our classification task, we generated ground-state wavefunctions via exact diagonalization of the Hamiltonian in Eq.~\eqref{eq:energy_2} across a range of these parameter settings. We then mapped the Trivial phase to $\ket{00}$, the symmetry-broken (SB) phase to $\ket{01}$, and the symmetry-protected topological (SPT) phase to $\ket{10}$. An additional label $\ket{11}$ was reserved to capture failure cases or out-of-distribution classifications.

For the AIQT, we implemented the parameterized QFT instance described above, with a single learnable global parameter $\theta$ that controls the interpolation between the Hadamard transform and the full QFT. During training, $\theta$ was initialized to $2\pi$ (corresponding to the standard QFT) and was jointly optimized alongside the QNN parameters.

The hardware-efficient QNN used in this study consists of repeated two-qubit entangling layers (Fig.~\ref{fig1}(c)). Each layer is implemented as an exponential of a parameterized Hermitian operator acting on qubit pairs. Specifically, at layer $\ell$, we define the two-qubit unitaries as:
\begin{equation}
  U_\ell = \exp(-i H_\ell),
\end{equation}
where the Hermitian generator $H_\ell$ is expressed in terms of the complete, orthonormal basis of $4\times 4$ Gell-Mann matrices $\{G_i\}$ \cite{bertlmann_bloch_2008}. These $G_i$ form the 15 traceless, Hermitian generators of SU(4) and constitute an orthonormal basis under the Hilbert-Schmidt inner product. Using this basis, $H_\ell$ is written as:
\begin{equation}
  H_\ell = \sum_{i=1}^{15} \phi_{\ell,i}\,G_i.
\end{equation}
Here, $\{\phi_{\ell,i}\}$ are real-valued trainable parameters for layer $\ell$, shared across all two-qubit pairs within the layer (Fig.~\ref{fig1}(c)).

The ground-state wavefunction was amplitude-encoded as the input state $|\psi_{\mathrm{in}}\rangle$. The AIQT and QNN layers are then applied sequentially to produce the output state $|\psi_{\mathrm{out}}\rangle$ (Eq.~\ref{eq:in_out}). The final prediction is obtained by measuring a set of $N_{\mathrm{tq}}$ target qubits in the computational basis. Here, $N_{\mathrm{tq}}$ denotes the number of target qubits, which are the qubits measured at the end of the circuit to produce the model’s output in the computational basis. The probability of obtaining outcome $i$ is given by:
\begin{gather}
    f_{i}(|\psi_{\mathrm{in}} \rangle; \theta, \boldsymbol{\phi})
    =
    \langle \psi_{\mathrm{out}} |P_i| \psi_{\mathrm{out}} \rangle,
\end{gather}
where $\{P_i\}_{i=0,\ldots, 2^{N_\mathrm{tq}}-1}$ is the set of projection operators onto the computational basis states $|i\rangle$ of the target qubits. For the classification of quantum phases described above, we set $N_{\mathrm{tq}} = 2$.

The model's objective is to map an input state $|\psi_{\mathrm{in}}\rangle$ to a corresponding one-hot encoded label vector $\boldsymbol{y}$ representing the quantum phase class. Given a training dataset of $M$ examples, denoted $\{ (|\psi_{\mathrm{in}}^{\alpha}\rangle, \boldsymbol{y}^{\alpha}) \}_{\alpha=1,\ldots,M}$, the model parameters are optimized by minimizing a loss function. In this study, we utilize the cross-entropy (log-loss) function:
\begin{equation}
    \label{eq:loss_function}
    L(\theta, \boldsymbol{\phi})
    =
    - \frac{1}{M}
    \sum_{\alpha=1}^{M}
    \sum_{i=0}^{2^{N_{\mathrm{tq}}} - 1}
    y_i^{\alpha}
    \log f_{i}(|\psi_{\mathrm{in}}^{\alpha}\rangle; \theta, \boldsymbol{\phi}).
\end{equation}

During training, the model predicts a probability distribution over the $2^{N_{\mathrm{tq}}}$ computational basis states by measuring the final $N_{\mathrm{tq}}$ target qubits. Both the global AIQT parameter $\theta$ and the local QNN parameters $\boldsymbol{\phi}$ are jointly optimized to minimize the loss function in Eq.~\eqref{eq:loss_function} using the Adam optimizer with a default learning rate of $5 \times 10^{-3}$. Training is performed with a batch size of 32. Models are trained and evaluated using standard 70:30 stratified train-test splits to maintain balanced class distributions across phases, with a total of 900 data points evenly divided among the three classes. All models and training were implemented using the PyTorch framework.

\begin{figure}[ht]
    \centering
    \includegraphics[width=0.48\textwidth]{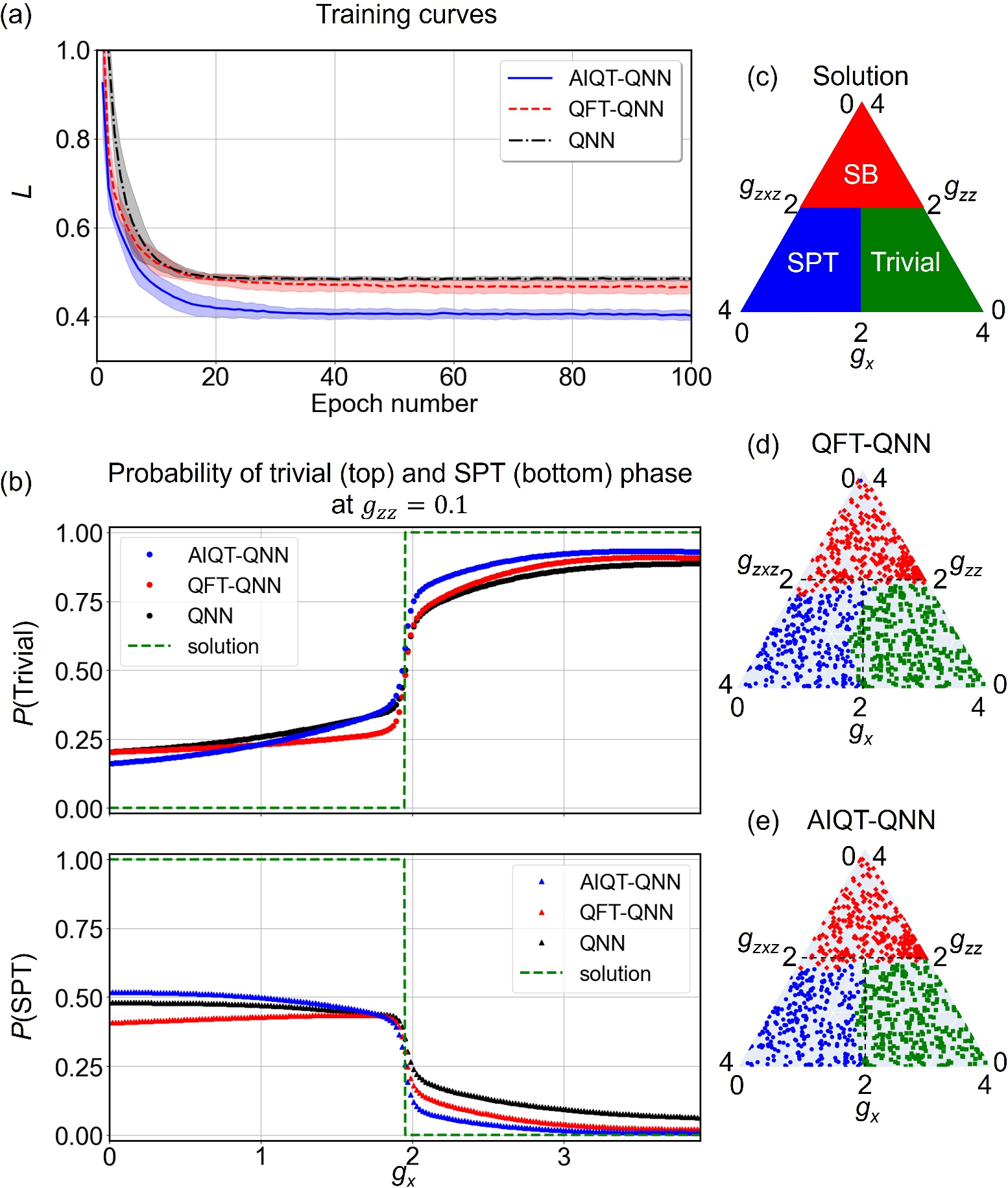}
    \caption{(a) The training curves of AIQT-QNN (blue), QFT-QNN (red), and QNN (black) for a 10-qubit system. $L$ denotes the loss value as defined by the loss function in Eq.~\ref{eq:loss_function}. The shaded region represents the standard deviation for 10 random initial parameters. Across all models, 3 layers of QNN were used. (b) The probability of being in the trivial phase (\textit{P}(Trivial)) (top)  and SPT (\textit{P}(SPT)) (bottom) phase at $g_{zz}=0.1$ of AIQT-QNN (blue), QFT-QNN (red), and QNN (black) after 100 training epochs. (c) The phase diagram of the Hamiltonian described in Eq. (8) as reported in literature \cite{PhysRevB.96.165124, PhysRevLett.120.057001, smith_crossing_2022}. The classification of the quantum phases using QFT-QNN (d) and AIQT-QNN (e). Here, $g_{zxz}$, $g_{zz}$, and $g_{x}$ are the parameters of the Hamiltonian described in Eq.~\ref{eq:energy_2}.}
    \label{fig2}
\end{figure}

\section{Results and Discussion}
\subsection*{AIQT using Parametrized QFT}

The performance of AIQT-QNN, QFT-QNN, and QNN models, each with three QNN layers, was compared on a 10-spin quantum phase classification task. Figure~\ref{fig2}(a) shows the training loss (\textit{L}) curves over 100 epochs. Both QFT-QNN and AIQT-QNN outperform the baseline QNN by converging to lower loss values, with AIQT-QNN achieving the lowest final loss, indicating superior training efficiency and optimization.

Figure~\ref{fig2}(b) shows the predicted class probabilities at fixed $g_{zz} = 0.1$, with the top panel depicting the probability of the Trivial phase (\textit{P}(Trivial)) and the bottom panel the SPT phase (\textit{P}(SPT)). The analytical solution boundary is indicated by a dashed green line. Here, AIQT-QNN exhibits the sharpest transitions at the phase boundary, most closely matching the analytical solution for both phases. Compared to QFT-QNN and QNN models, the AIQT-QNN consistently demonstrates steeper slopes and reduced probability leakage across the transition, highlighting its enhanced ability to distinguish between phases.

\begin{figure}[ht]
    \centering
    \includegraphics[width=0.45\textwidth]{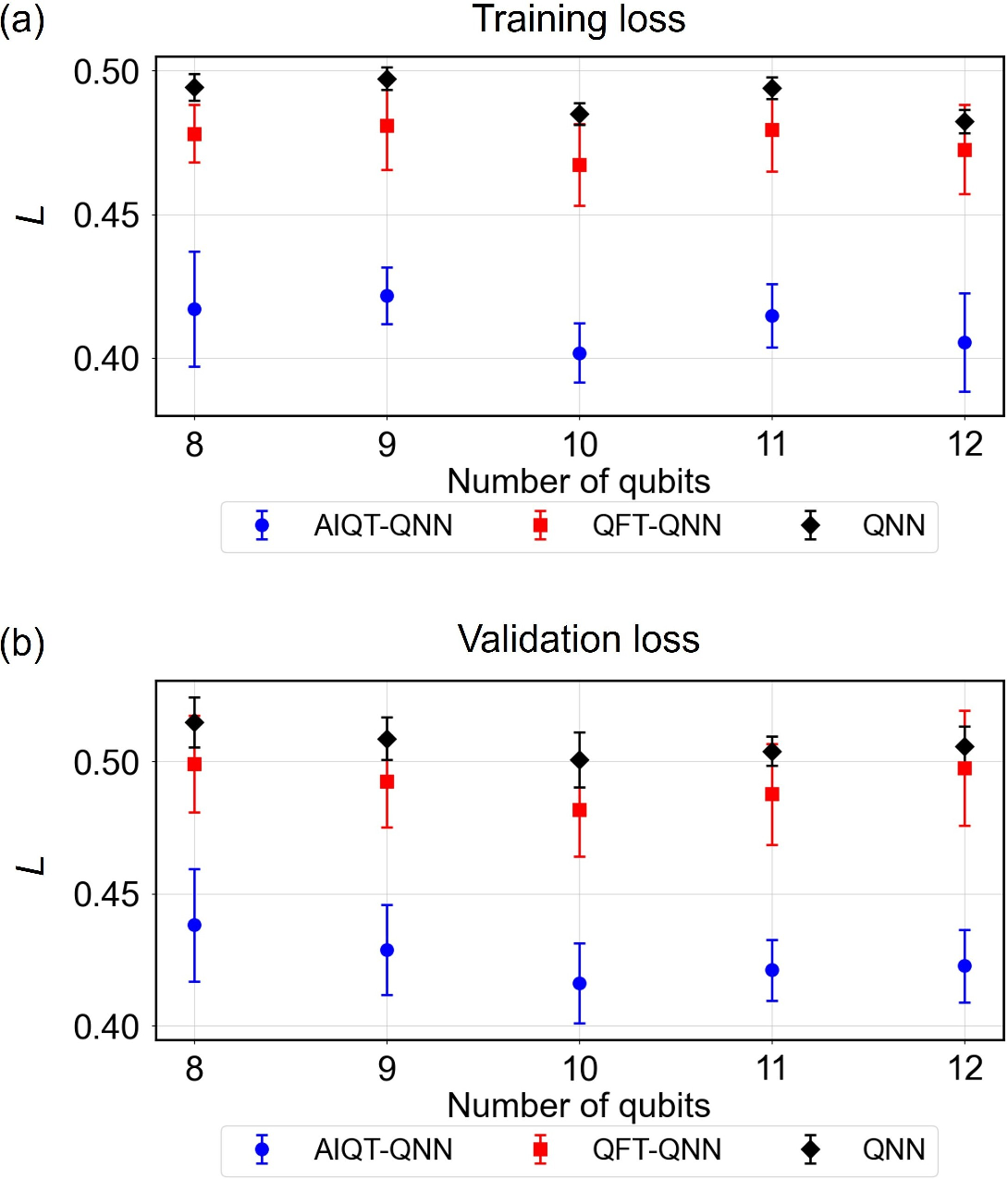}
    \caption{Training loss (a) and validation loss (b) as a function of the number of qubits for AIQT-QNN (blue), QFT-QNN (red), and baseline QNN (black) after 100 training epochs. The markers and whiskers show the average and standard deviation values derived from 10 random initial parameters.
    }
    \label{fig_n_qubits}
\end{figure}

Figure ~\ref{fig2}(c) depicts the ground-truth phase diagram in the $(g_x, g_{zz}, g_{zxz})$ space, delineating SPT, SB, and Trivial regions. Figures ~\ref{fig2}(d) and ~\ref{fig2}(e) show the predicted phase classifications for QFT-QNN and AIQT-QNN, respectively. While both models recover the overall phase structure, the AIQT-QNN model achieves better alignment with analytical boundaries, reducing misclassifications and yielding clearer phase separation, particularly near the SPT-Trivial and SPT-SB boundaries. These consistent improvements highlight the effectiveness of the AIQT module in enhancing QNN classification performance.

Figure~\ref{fig_n_qubits} shows the training and validation loss performance of QNN-based models as a function of the number of qubits. Figure~\ref{fig_n_qubits}(a) presents the training loss, while Fig.~\ref{fig_n_qubits}(b) shows the corresponding validation loss. In both cases, AIQT-QNN (blue) achieves the lowest loss values across all number of qubits, outperforming both QFT-QNN (red) and the baseline QNN (black).

We argue that the observed advantage of AIQT arises from two fundamental mechanisms. First, AIQT introduces a small number of learnable parameters that enable adaptive feature extraction from the input data. This design allows the model to dynamically adjust the preprocessing stage in a way that minimizes the overall loss function. Unlike fixed unitary transforms such as the QFT, AIQT supports task-specific preprocessing and thereby improves adaptability. Second, AIQT offers explicit, learnable control over the amount of entanglement injected into the input state of the QNN. This capability is particularly important in hardware-efficient QNNs, where the success of gradient-based optimization is known to depend strongly on the entanglement structure of the initial state. Motivated by recent findings \cite{Leone2024practicalusefulness} that highlight a “Goldilocks zone” of entanglement, balancing expressivity and trainability, AIQT acts as an adaptive preprocessing layer that guides the input state toward this favorable regime. Together, these two mechanisms position AIQT as an alternative to fixed global transforms or deep variational encoders, offering a principled route toward more efficient and scalable quantum learning models.

\subsection*{AIQT using Time Evolution of Transverse-field Ising Hamiltonian}
In addition to this Fourier-domain example, we also introduce an instance of the AIQT framework based on the time evolution of the transverse-field Ising model (TFIM). In this case, the AIQT is defined as a parameterized time-evolution operator of the form
\begin{equation}
    \label{eq:AIQT_2}
    U_{\mathrm{AIQT}}(\theta, J, g) = e^{-i \theta \, H_{\text{TFIM}}(J, g)},
\end{equation}
where \(\theta\) is a learnable global time parameter and \(J, g\) are learnable physical coupling strengths. The TFIM Hamiltonian is given by
\begin{equation}
    H_{\text{TFIM}}(J, g) = -J \sum_{i} Z_i Z_{i+1} - g \sum_{i} X_i.
\end{equation}

To make this evolution tractable and differentiable in simulation, we approximate the time-evolution operator using a multi-step first-order Trotter-Suzuki decomposition. Specifically, the total evolution time \(\theta\) is divided into \(n_{\text{steps}}\) intervals of duration \(\delta\theta = \theta / n_{\text{steps}}\). For each Trotter step, the unitary is factorized into:
\begin{itemize}
    \item Single-qubit \(X\)-rotations of angle \(-2g\,\delta\theta\) applied independently on each qubit, representing the transverse field term.
    \item The \(ZZ\) couplings are realized using a CNOT-RZ-CNOT pattern, with the RZ gate having the angle of \(-2J\,\delta\theta\).
\end{itemize}
The complete Time Evolution (TE)-based \(U_{\mathrm{AIQT}}\) is then constructed by applying this Trotter-step unitary \(n_{\text{steps}}\) times in sequence. In our experiments, the number of Trotter steps is fixed at  \(n_{\text{steps}}=10\). By optimizing $\theta$, $J$, and $g$ parameters, the model may discover the coupling regime and timescale at which quantum phase features become most discriminative for the classification task. In the limit \(\theta \to 0\), the unitary naturally recovers the Identity operator (\(U_A = I\)), while for non-zero \(\theta\) it generates entanglement patterns governed by the learned TFIM Hamiltonian, while remaining efficiently implementable on quantum hardware using standard Trotterization techniques.

\begin{figure}[t]
    \centering
    \includegraphics[width=0.45\textwidth]{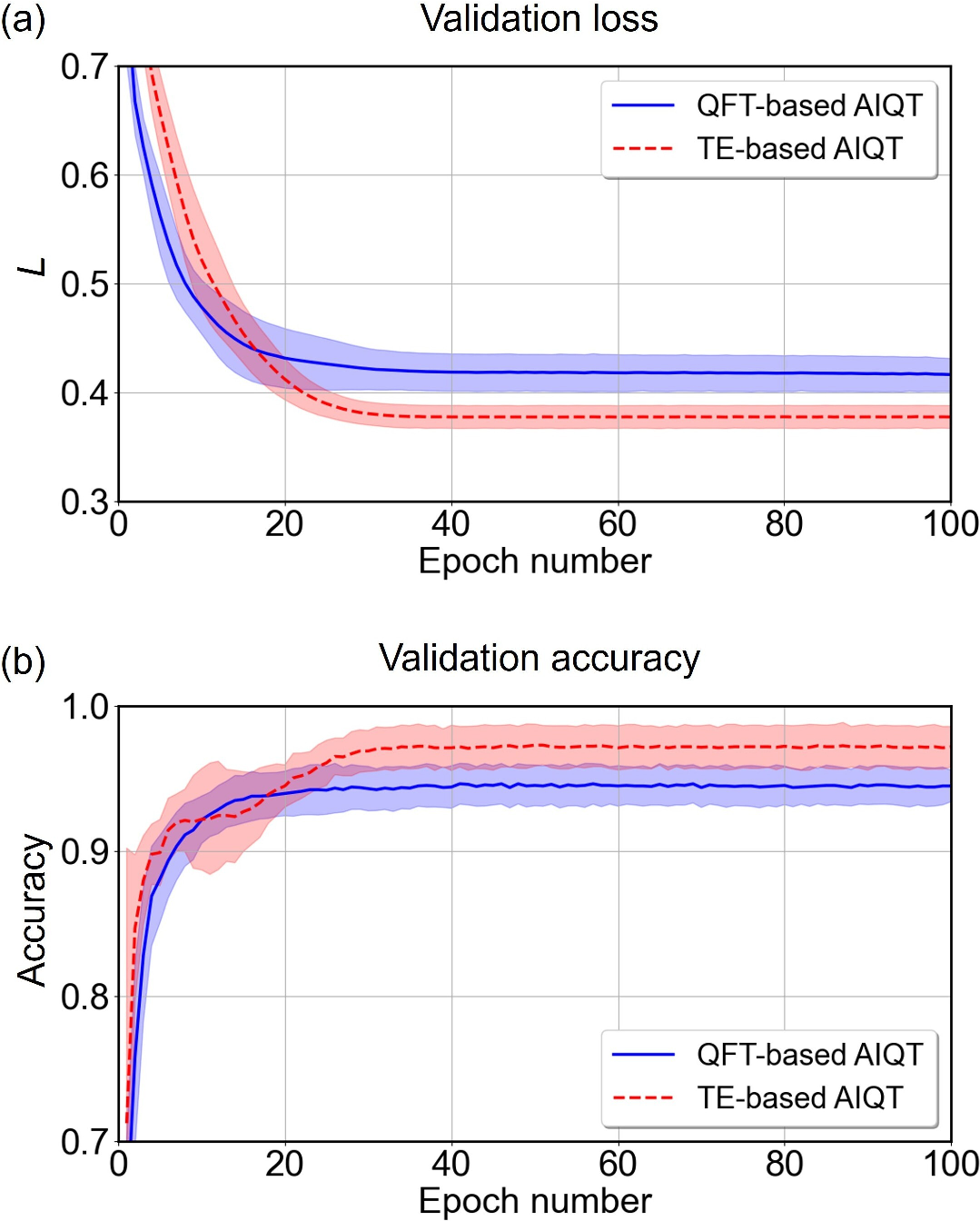}
    \caption{Validation performance in loss (a) and accuracy (b) for models using two different AIQT variants: QFT-based AIQT (blue) and Time Evolution (TE)-based AIQT with TFIM Hamiltonian (red) for a 10-qubit system, evaluated over 100 training epochs. Shaded regions indicate standard deviation across 10 training runs. 
    }
    \label{fig_tfim}
\end{figure}

Figure~\ref{fig_tfim} shows the validation performance of models using two different AIQT definitions: QFT-based AIQT (blue) and Time Evolution (TE)-based AIQT (red), evaluated over 100 training epochs. As shown in Fig.~\ref{fig_tfim}(a), the TE-based AIQT model consistently achieves lower final loss values than the QFT-based variant, indicating improved generalization performance. Figure~\ref{fig_tfim}(b) displays the corresponding validation accuracy curves. Both models reach high classification accuracy exceeding 90\%, with the TE-based AIQT model reaching slightly higher final accuracy. These results demonstrate that, for this task, the TE-based AIQT leads to more efficient training and improved performance compared to the QFT-based approach.

This outcome illustrates how the effectiveness of an AIQT variant depends on the choice of instantiation: QFT-based AIQT is well-matched to global symmetries, whereas the TE-based approach reflects local interaction structures. Beyond feature extraction, each variant also introduces a different entanglement profile into the input state, which can significantly influence optimization dynamics and model expressivity. Together, this highlights the flexibility of AIQT in adapting to different structural and physical properties of quantum learning tasks.


\section{Conclusion}
In this work, we introduced the Adaptive Interpolating Quantum Transform (AIQT), a quantum-native framework that defines a trainable unitary operator interpolating between known quantum transforms with minimal global parameters. This approach provides an explicit inductive bias while allowing flexible, adaptive control over input-state entanglement and information mixing in variational quantum circuits.

We tested AIQT-enhanced models on a quantum phase classification task, showing that they outperform both baseline QNNs and models using fixed QFT layers. This improved performance reflects the AIQT layer's ability to adaptively control the amount and structure of entanglement and information mixing introduced into the input state, enabling the model to discover task-relevant representations while avoiding over-scrambling that hinders optimization.

\begin{acknowledgments}
This work was supported by JSPS KAKENHI under Grant-in-Aid for Early-Career Scientists No. JP24K16985 and JSPS KAKENHI under Grant-in-Aid for Transformative Research Areas No. JP22H05114. This study was partially carried out using the facilities of the Supercomputer Center, the Institute for Solid State Physics, the University of Tokyo and the TSUBAME4.0 supercomputer at the Institute of Science Tokyo. This work was partially supported by the Center of Innovations for Sustainable Quantum AI (JST Grant Number JPMJPF2221). R.K. was supported by JSPS KAKENHI (Grants No. JP24H00973 and No. JP25K07157). This work was supported by Japan Science and Technology Agency (JST) as part of Adopting Sustainable Partnerships for Innovative Research Ecosystem (ASPIRE), Grant Number JPMJAP24C1. The author acknowledges the contributions and discussions provided by the members of Quemix Inc.

\end{acknowledgments}
\bibliography{manuscript}

\begin{thebibliography}{41}%
\makeatletter
\providecommand \@ifxundefined [1]{%
 \@ifx{#1\undefined}
}%
\providecommand \@ifnum [1]{%
 \ifnum #1\expandafter \@firstoftwo
 \else \expandafter \@secondoftwo
 \fi
}%
\providecommand \@ifx [1]{%
 \ifx #1\expandafter \@firstoftwo
 \else \expandafter \@secondoftwo
 \fi
}%
\providecommand \natexlab [1]{#1}%
\providecommand \enquote  [1]{``#1''}%
\providecommand \bibnamefont  [1]{#1}%
\providecommand \bibfnamefont [1]{#1}%
\providecommand \citenamefont [1]{#1}%
\providecommand \href@noop [0]{\@secondoftwo}%
\providecommand \href [0]{\begingroup \@sanitize@url \@href}%
\providecommand \@href[1]{\@@startlink{#1}\@@href}%
\providecommand \@@href[1]{\endgroup#1\@@endlink}%
\providecommand \@sanitize@url [0]{\catcode `\\12\catcode `\$12\catcode `\&12\catcode `\#12\catcode `\^12\catcode `\_12\catcode `\%12\relax}%
\providecommand \@@startlink[1]{}%
\providecommand \@@endlink[0]{}%
\providecommand \url  [0]{\begingroup\@sanitize@url \@url }%
\providecommand \@url [1]{\endgroup\@href {#1}{\urlprefix }}%
\providecommand \urlprefix  [0]{URL }%
\providecommand \Eprint [0]{\href }%
\providecommand \doibase [0]{https://doi.org/}%
\providecommand \selectlanguage [0]{\@gobble}%
\providecommand \bibinfo  [0]{\@secondoftwo}%
\providecommand \bibfield  [0]{\@secondoftwo}%
\providecommand \translation [1]{[#1]}%
\providecommand \BibitemOpen [0]{}%
\providecommand \bibitemStop [0]{}%
\providecommand \bibitemNoStop [0]{.\EOS\space}%
\providecommand \EOS [0]{\spacefactor3000\relax}%
\providecommand \BibitemShut  [1]{\csname bibitem#1\endcsname}%
\let\auto@bib@innerbib\@empty
\bibitem [{\citenamefont {He}\ \emph {et~al.}(2016)\citenamefont {He}, \citenamefont {Zhang}, \citenamefont {Ren},\ and\ \citenamefont {Sun}}]{resnet50}%
  \BibitemOpen
  \bibfield  {author} {\bibinfo {author} {\bibfnamefont {K.}~\bibnamefont {He}}, \bibinfo {author} {\bibfnamefont {X.}~\bibnamefont {Zhang}}, \bibinfo {author} {\bibfnamefont {S.}~\bibnamefont {Ren}},\ and\ \bibinfo {author} {\bibfnamefont {J.}~\bibnamefont {Sun}},\ }\bibfield  {title} {\bibinfo {title} {Deep residual learning for image recognition},\ }in\ \href {https://doi.org/10.1109/CVPR.2016.90} {\emph {\bibinfo {booktitle} {2016 IEEE Conference on Computer Vision and Pattern Recognition (CVPR)}}}\ (\bibinfo {year} {2016})\ pp.\ \bibinfo {pages} {770--778}\BibitemShut {NoStop}%
\bibitem [{\citenamefont {Vaswani}\ \emph {et~al.}(2023)\citenamefont {Vaswani}, \citenamefont {Shazeer}, \citenamefont {Parmar}, \citenamefont {Uszkoreit}, \citenamefont {Jones}, \citenamefont {Gomez}, \citenamefont {Kaiser},\ and\ \citenamefont {Polosukhin}}]{vaswani2023}%
  \BibitemOpen
  \bibfield  {author} {\bibinfo {author} {\bibfnamefont {A.}~\bibnamefont {Vaswani}}, \bibinfo {author} {\bibfnamefont {N.}~\bibnamefont {Shazeer}}, \bibinfo {author} {\bibfnamefont {N.}~\bibnamefont {Parmar}}, \bibinfo {author} {\bibfnamefont {J.}~\bibnamefont {Uszkoreit}}, \bibinfo {author} {\bibfnamefont {L.}~\bibnamefont {Jones}}, \bibinfo {author} {\bibfnamefont {A.~N.}\ \bibnamefont {Gomez}}, \bibinfo {author} {\bibfnamefont {L.}~\bibnamefont {Kaiser}},\ and\ \bibinfo {author} {\bibfnamefont {I.}~\bibnamefont {Polosukhin}},\ }\href {https://arxiv.org/abs/1706.03762} {\bibinfo {title} {Attention is all you need}} (\bibinfo {year} {2023}),\ \Eprint {https://arxiv.org/abs/1706.03762} {arXiv:1706.03762 [cs.CL]} \BibitemShut {NoStop}%
\bibitem [{\citenamefont {Brown}\ \emph {et~al.}(2020)\citenamefont {Brown}, \citenamefont {Mann}, \citenamefont {Ryder}, \citenamefont {Subbiah}, \citenamefont {Kaplan}, \citenamefont {Dhariwal}, \citenamefont {Neelakantan}, \citenamefont {Shyam}, \citenamefont {Sastry}, \citenamefont {Askell}, \citenamefont {Agarwal}, \citenamefont {Herbert-Voss}, \citenamefont {Krueger}, \citenamefont {Henighan}, \citenamefont {Child}, \citenamefont {Ramesh}, \citenamefont {Ziegler}, \citenamefont {Wu}, \citenamefont {Winter}, \citenamefont {Hesse}, \citenamefont {Chen}, \citenamefont {Sigler}, \citenamefont {Litwin}, \citenamefont {Gray}, \citenamefont {Chess}, \citenamefont {Clark}, \citenamefont {Berner}, \citenamefont {McCandlish}, \citenamefont {Radford}, \citenamefont {Sutskever},\ and\ \citenamefont {Amodei}}]{brown2020}%
  \BibitemOpen
  \bibfield  {author} {\bibinfo {author} {\bibfnamefont {T.~B.}\ \bibnamefont {Brown}}, \bibinfo {author} {\bibfnamefont {B.}~\bibnamefont {Mann}}, \bibinfo {author} {\bibfnamefont {N.}~\bibnamefont {Ryder}}, \bibinfo {author} {\bibfnamefont {M.}~\bibnamefont {Subbiah}}, \bibinfo {author} {\bibfnamefont {J.}~\bibnamefont {Kaplan}}, \bibinfo {author} {\bibfnamefont {P.}~\bibnamefont {Dhariwal}}, \bibinfo {author} {\bibfnamefont {A.}~\bibnamefont {Neelakantan}}, \bibinfo {author} {\bibfnamefont {P.}~\bibnamefont {Shyam}}, \bibinfo {author} {\bibfnamefont {G.}~\bibnamefont {Sastry}}, \bibinfo {author} {\bibfnamefont {A.}~\bibnamefont {Askell}}, \bibinfo {author} {\bibfnamefont {S.}~\bibnamefont {Agarwal}}, \bibinfo {author} {\bibfnamefont {A.}~\bibnamefont {Herbert-Voss}}, \bibinfo {author} {\bibfnamefont {G.}~\bibnamefont {Krueger}}, \bibinfo {author} {\bibfnamefont {T.}~\bibnamefont {Henighan}}, \bibinfo {author} {\bibfnamefont {R.}~\bibnamefont {Child}}, \bibinfo {author} {\bibfnamefont {A.}~\bibnamefont
  {Ramesh}}, \bibinfo {author} {\bibfnamefont {D.~M.}\ \bibnamefont {Ziegler}}, \bibinfo {author} {\bibfnamefont {J.}~\bibnamefont {Wu}}, \bibinfo {author} {\bibfnamefont {C.}~\bibnamefont {Winter}}, \bibinfo {author} {\bibfnamefont {C.}~\bibnamefont {Hesse}}, \bibinfo {author} {\bibfnamefont {M.}~\bibnamefont {Chen}}, \bibinfo {author} {\bibfnamefont {E.}~\bibnamefont {Sigler}}, \bibinfo {author} {\bibfnamefont {M.}~\bibnamefont {Litwin}}, \bibinfo {author} {\bibfnamefont {S.}~\bibnamefont {Gray}}, \bibinfo {author} {\bibfnamefont {B.}~\bibnamefont {Chess}}, \bibinfo {author} {\bibfnamefont {J.}~\bibnamefont {Clark}}, \bibinfo {author} {\bibfnamefont {C.}~\bibnamefont {Berner}}, \bibinfo {author} {\bibfnamefont {S.}~\bibnamefont {McCandlish}}, \bibinfo {author} {\bibfnamefont {A.}~\bibnamefont {Radford}}, \bibinfo {author} {\bibfnamefont {I.}~\bibnamefont {Sutskever}},\ and\ \bibinfo {author} {\bibfnamefont {D.}~\bibnamefont {Amodei}},\ }\href {https://arxiv.org/abs/2005.14165} {\bibinfo {title} {Language
  models are few-shot learners}} (\bibinfo {year} {2020}),\ \Eprint {https://arxiv.org/abs/2005.14165} {arXiv:2005.14165 [cs.CL]} \BibitemShut {NoStop}%
\bibitem [{\citenamefont {Kaplan}\ \emph {et~al.}(2020)\citenamefont {Kaplan}, \citenamefont {McCandlish}, \citenamefont {Henighan}, \citenamefont {Brown}, \citenamefont {Chess}, \citenamefont {Child}, \citenamefont {Gray}, \citenamefont {Radford}, \citenamefont {Wu},\ and\ \citenamefont {Amodei}}]{kaplan2020}%
  \BibitemOpen
  \bibfield  {author} {\bibinfo {author} {\bibfnamefont {J.}~\bibnamefont {Kaplan}}, \bibinfo {author} {\bibfnamefont {S.}~\bibnamefont {McCandlish}}, \bibinfo {author} {\bibfnamefont {T.}~\bibnamefont {Henighan}}, \bibinfo {author} {\bibfnamefont {T.~B.}\ \bibnamefont {Brown}}, \bibinfo {author} {\bibfnamefont {B.}~\bibnamefont {Chess}}, \bibinfo {author} {\bibfnamefont {R.}~\bibnamefont {Child}}, \bibinfo {author} {\bibfnamefont {S.}~\bibnamefont {Gray}}, \bibinfo {author} {\bibfnamefont {A.}~\bibnamefont {Radford}}, \bibinfo {author} {\bibfnamefont {J.}~\bibnamefont {Wu}},\ and\ \bibinfo {author} {\bibfnamefont {D.}~\bibnamefont {Amodei}},\ }\href {https://arxiv.org/abs/2001.08361} {\bibinfo {title} {Scaling laws for neural language models}} (\bibinfo {year} {2020}),\ \Eprint {https://arxiv.org/abs/2001.08361} {arXiv:2001.08361 [cs.LG]} \BibitemShut {NoStop}%
\bibitem [{\citenamefont {Jassim}\ and\ \citenamefont {Harte}(2022)}]{https://doi.org/10.1049/sil2.12109}%
  \BibitemOpen
  \bibfield  {author} {\bibinfo {author} {\bibfnamefont {W.~A.}\ \bibnamefont {Jassim}}\ and\ \bibinfo {author} {\bibfnamefont {N.}~\bibnamefont {Harte}},\ }\bibfield  {title} {\bibinfo {title} {Comparison of discrete transforms for deep-neural-networks-based speech enhancement},\ }\href {https://doi.org/https://doi.org/10.1049/sil2.12109} {\bibfield  {journal} {\bibinfo  {journal} {IET Signal Processing}\ }\textbf {\bibinfo {volume} {16}},\ \bibinfo {pages} {438} (\bibinfo {year} {2022})}\BibitemShut {NoStop}%
\bibitem [{\citenamefont {Yi}\ \emph {et~al.}(2025)\citenamefont {Yi}, \citenamefont {Zhang}, \citenamefont {Fan}, \citenamefont {Cao}, \citenamefont {Wang}, \citenamefont {Long}, \citenamefont {Hu}, \citenamefont {He}, \citenamefont {Wen},\ and\ \citenamefont {Xiong}}]{Yi2023ASO}%
  \BibitemOpen
  \bibfield  {author} {\bibinfo {author} {\bibfnamefont {K.}~\bibnamefont {Yi}}, \bibinfo {author} {\bibfnamefont {Q.}~\bibnamefont {Zhang}}, \bibinfo {author} {\bibfnamefont {W.}~\bibnamefont {Fan}}, \bibinfo {author} {\bibfnamefont {L.}~\bibnamefont {Cao}}, \bibinfo {author} {\bibfnamefont {S.}~\bibnamefont {Wang}}, \bibinfo {author} {\bibfnamefont {G.}~\bibnamefont {Long}}, \bibinfo {author} {\bibfnamefont {L.}~\bibnamefont {Hu}}, \bibinfo {author} {\bibfnamefont {H.}~\bibnamefont {He}}, \bibinfo {author} {\bibfnamefont {Q.}~\bibnamefont {Wen}},\ and\ \bibinfo {author} {\bibfnamefont {H.}~\bibnamefont {Xiong}},\ }\href {https://arxiv.org/abs/2302.02173} {\bibinfo {title} {A survey on deep learning based time series analysis with frequency transformation}} (\bibinfo {year} {2025}),\ \Eprint {https://arxiv.org/abs/2302.02173} {arXiv:2302.02173 [cs.LG]} \BibitemShut {NoStop}%
\bibitem [{\citenamefont {Guibas}\ \emph {et~al.}(2022)\citenamefont {Guibas}, \citenamefont {Mardani}, \citenamefont {Li}, \citenamefont {Tao}, \citenamefont {Anandkumar},\ and\ \citenamefont {Catanzaro}}]{guibas2022}%
  \BibitemOpen
  \bibfield  {author} {\bibinfo {author} {\bibfnamefont {J.}~\bibnamefont {Guibas}}, \bibinfo {author} {\bibfnamefont {M.}~\bibnamefont {Mardani}}, \bibinfo {author} {\bibfnamefont {Z.}~\bibnamefont {Li}}, \bibinfo {author} {\bibfnamefont {A.}~\bibnamefont {Tao}}, \bibinfo {author} {\bibfnamefont {A.}~\bibnamefont {Anandkumar}},\ and\ \bibinfo {author} {\bibfnamefont {B.}~\bibnamefont {Catanzaro}},\ }\href {https://arxiv.org/abs/2111.13587} {\bibinfo {title} {Adaptive fourier neural operators: Efficient token mixers for transformers}} (\bibinfo {year} {2022}),\ \Eprint {https://arxiv.org/abs/2111.13587} {arXiv:2111.13587 [cs.CV]} \BibitemShut {NoStop}%
\bibitem [{\citenamefont {Tancik}\ \emph {et~al.}(2020)\citenamefont {Tancik}, \citenamefont {Srinivasan}, \citenamefont {Mildenhall}, \citenamefont {Fridovich-Keil}, \citenamefont {Raghavan}, \citenamefont {Singhal}, \citenamefont {Ramamoorthi}, \citenamefont {Barron},\ and\ \citenamefont {Ng}}]{tancik2020fourier}%
  \BibitemOpen
  \bibfield  {author} {\bibinfo {author} {\bibfnamefont {M.}~\bibnamefont {Tancik}}, \bibinfo {author} {\bibfnamefont {P.}~\bibnamefont {Srinivasan}}, \bibinfo {author} {\bibfnamefont {B.}~\bibnamefont {Mildenhall}}, \bibinfo {author} {\bibfnamefont {S.}~\bibnamefont {Fridovich-Keil}}, \bibinfo {author} {\bibfnamefont {N.}~\bibnamefont {Raghavan}}, \bibinfo {author} {\bibfnamefont {U.}~\bibnamefont {Singhal}}, \bibinfo {author} {\bibfnamefont {R.}~\bibnamefont {Ramamoorthi}}, \bibinfo {author} {\bibfnamefont {J.}~\bibnamefont {Barron}},\ and\ \bibinfo {author} {\bibfnamefont {R.}~\bibnamefont {Ng}},\ }\bibfield  {title} {\bibinfo {title} {Fourier features let networks learn high frequency functions in low dimensional domains},\ }\href@noop {} {\bibfield  {journal} {\bibinfo  {journal} {Advances in neural information processing systems}\ }\textbf {\bibinfo {volume} {33}},\ \bibinfo {pages} {7537} (\bibinfo {year} {2020})}\BibitemShut {NoStop}%
\bibitem [{\citenamefont {Lou}\ \emph {et~al.}(2021)\citenamefont {Lou}, \citenamefont {Park}, \citenamefont {Ramezanali},\ and\ \citenamefont {Tang}}]{lou2021fnetarmixingtokensautoregressive}%
  \BibitemOpen
  \bibfield  {author} {\bibinfo {author} {\bibfnamefont {T.}~\bibnamefont {Lou}}, \bibinfo {author} {\bibfnamefont {M.}~\bibnamefont {Park}}, \bibinfo {author} {\bibfnamefont {M.}~\bibnamefont {Ramezanali}},\ and\ \bibinfo {author} {\bibfnamefont {V.}~\bibnamefont {Tang}},\ }\href {https://arxiv.org/abs/2107.10932} {\bibinfo {title} {Fnetar: Mixing tokens with autoregressive fourier transforms}} (\bibinfo {year} {2021}),\ \Eprint {https://arxiv.org/abs/2107.10932} {arXiv:2107.10932 [cs.CL]} \BibitemShut {NoStop}%
\bibitem [{\citenamefont {Wang}\ \emph {et~al.}(2024)\citenamefont {Wang}, \citenamefont {Wu}, \citenamefont {Shi}, \citenamefont {Hu}, \citenamefont {Luo}, \citenamefont {Ma}, \citenamefont {Zhang},\ and\ \citenamefont {ZHOU}}]{wang2024timemixer}%
  \BibitemOpen
  \bibfield  {author} {\bibinfo {author} {\bibfnamefont {S.}~\bibnamefont {Wang}}, \bibinfo {author} {\bibfnamefont {H.}~\bibnamefont {Wu}}, \bibinfo {author} {\bibfnamefont {X.}~\bibnamefont {Shi}}, \bibinfo {author} {\bibfnamefont {T.}~\bibnamefont {Hu}}, \bibinfo {author} {\bibfnamefont {H.}~\bibnamefont {Luo}}, \bibinfo {author} {\bibfnamefont {L.}~\bibnamefont {Ma}}, \bibinfo {author} {\bibfnamefont {J.~Y.}\ \bibnamefont {Zhang}},\ and\ \bibinfo {author} {\bibfnamefont {J.}~\bibnamefont {ZHOU}},\ }\bibfield  {title} {\bibinfo {title} {Timemixer: Decomposable multiscale mixing for time series forecasting},\ }in\ \href {https://openreview.net/forum?id=7oLshfEIC2} {\emph {\bibinfo {booktitle} {The Twelfth International Conference on Learning Representations}}}\ (\bibinfo {year} {2024})\BibitemShut {NoStop}%
\bibitem [{\citenamefont {Xu}\ \emph {et~al.}(2020)\citenamefont {Xu}, \citenamefont {Qin}, \citenamefont {Sun}, \citenamefont {Wang}, \citenamefont {Chen},\ and\ \citenamefont {Ren}}]{Xu2020}%
  \BibitemOpen
  \bibfield  {author} {\bibinfo {author} {\bibfnamefont {K.}~\bibnamefont {Xu}}, \bibinfo {author} {\bibfnamefont {M.}~\bibnamefont {Qin}}, \bibinfo {author} {\bibfnamefont {F.}~\bibnamefont {Sun}}, \bibinfo {author} {\bibfnamefont {Y.}~\bibnamefont {Wang}}, \bibinfo {author} {\bibfnamefont {Y.-K.}\ \bibnamefont {Chen}},\ and\ \bibinfo {author} {\bibfnamefont {F.}~\bibnamefont {Ren}},\ }\bibfield  {title} {\bibinfo {title} {Learning in the frequency domain},\ }in\ \href {https://doi.org/10.1109/CVPR42600.2020.00181} {\emph {\bibinfo {booktitle} {2020 IEEE/CVF Conference on Computer Vision and Pattern Recognition (CVPR)}}}\ (\bibinfo {year} {2020})\ pp.\ \bibinfo {pages} {1737--1746}\BibitemShut {NoStop}%
\bibitem [{\citenamefont {Ng}\ and\ \citenamefont {Beng Jin~Teoh}(2015)}]{7415375}%
  \BibitemOpen
  \bibfield  {author} {\bibinfo {author} {\bibfnamefont {C.~J.}\ \bibnamefont {Ng}}\ and\ \bibinfo {author} {\bibfnamefont {A.}~\bibnamefont {Beng Jin~Teoh}},\ }\bibfield  {title} {\bibinfo {title} {Dctnet: A simple learning-free approach for face recognition},\ }in\ \href {https://doi.org/10.1109/APSIPA.2015.7415375} {\emph {\bibinfo {booktitle} {2015 Asia-Pacific Signal and Information Processing Association Annual Summit and Conference (APSIPA)}}}\ (\bibinfo {year} {2015})\ pp.\ \bibinfo {pages} {761--768}\BibitemShut {NoStop}%
\bibitem [{\citenamefont {Karaoglu}\ and\ \citenamefont {Eksioglu}(2023)}]{Karaoglu2023}%
  \BibitemOpen
  \bibfield  {author} {\bibinfo {author} {\bibfnamefont {H.~H.}\ \bibnamefont {Karaoglu}}\ and\ \bibinfo {author} {\bibfnamefont {E.~M.}\ \bibnamefont {Eksioglu}},\ }\bibfield  {title} {\bibinfo {title} {Dctnet: deep shrinkage denoising via dct filterbanks},\ }\href {https://doi.org/10.1007/s11760-023-02593-0} {\bibfield  {journal} {\bibinfo  {journal} {Signal, Image and Video Processing}\ }\textbf {\bibinfo {volume} {17}},\ \bibinfo {pages} {3665–3676} (\bibinfo {year} {2023})}\BibitemShut {NoStop}%
\bibitem [{\citenamefont {Lee}\ and\ \citenamefont {Kim}(2024)}]{Lee2024}%
  \BibitemOpen
  \bibfield  {author} {\bibinfo {author} {\bibfnamefont {J.}~\bibnamefont {Lee}}\ and\ \bibinfo {author} {\bibfnamefont {H.}~\bibnamefont {Kim}},\ }\bibfield  {title} {\bibinfo {title} {Dct-vit: High-frequency pruned vision transformer with discrete cosine transform},\ }\href {https://doi.org/10.1109/access.2024.3410231} {\bibfield  {journal} {\bibinfo  {journal} {IEEE Access}\ }\textbf {\bibinfo {volume} {12}},\ \bibinfo {pages} {80386–80396} (\bibinfo {year} {2024})}\BibitemShut {NoStop}%
\bibitem [{\citenamefont {Su}\ \emph {et~al.}(2024)\citenamefont {Su}, \citenamefont {Cao}, \citenamefont {Zhao}, \citenamefont {Li}, \citenamefont {Wu}, \citenamefont {Han},\ and\ \citenamefont {Liu}}]{SU2024106139}%
  \BibitemOpen
  \bibfield  {author} {\bibinfo {author} {\bibfnamefont {K.}~\bibnamefont {Su}}, \bibinfo {author} {\bibfnamefont {L.}~\bibnamefont {Cao}}, \bibinfo {author} {\bibfnamefont {B.}~\bibnamefont {Zhao}}, \bibinfo {author} {\bibfnamefont {N.}~\bibnamefont {Li}}, \bibinfo {author} {\bibfnamefont {D.}~\bibnamefont {Wu}}, \bibinfo {author} {\bibfnamefont {X.}~\bibnamefont {Han}},\ and\ \bibinfo {author} {\bibfnamefont {Y.}~\bibnamefont {Liu}},\ }\bibfield  {title} {\bibinfo {title} {Dctvit: Discrete cosine transform meet vision transformers},\ }\href {https://doi.org/https://doi.org/10.1016/j.neunet.2024.106139} {\bibfield  {journal} {\bibinfo  {journal} {Neural Networks}\ }\textbf {\bibinfo {volume} {172}},\ \bibinfo {pages} {106139} (\bibinfo {year} {2024})}\BibitemShut {NoStop}%
\bibitem [{\citenamefont {Liu}\ \emph {et~al.}(2019)\citenamefont {Liu}, \citenamefont {Zhang}, \citenamefont {Lian},\ and\ \citenamefont {Zuo}}]{liu2019multi}%
  \BibitemOpen
  \bibfield  {author} {\bibinfo {author} {\bibfnamefont {P.}~\bibnamefont {Liu}}, \bibinfo {author} {\bibfnamefont {H.}~\bibnamefont {Zhang}}, \bibinfo {author} {\bibfnamefont {W.}~\bibnamefont {Lian}},\ and\ \bibinfo {author} {\bibfnamefont {W.}~\bibnamefont {Zuo}},\ }\bibfield  {title} {\bibinfo {title} {Multi-level wavelet convolutional neural networks},\ }\href {https://doi.org/10.1109/ACCESS.2019.2921451} {\bibfield  {journal} {\bibinfo  {journal} {IEEE Access}\ }\textbf {\bibinfo {volume} {7}},\ \bibinfo {pages} {74973} (\bibinfo {year} {2019})}\BibitemShut {NoStop}%
\bibitem [{\citenamefont {Yu}\ \emph {et~al.}(2020)\citenamefont {Yu}, \citenamefont {Derpanis},\ and\ \citenamefont {Brubaker}}]{WaveletFlow}%
  \BibitemOpen
  \bibfield  {author} {\bibinfo {author} {\bibfnamefont {J.~J.}\ \bibnamefont {Yu}}, \bibinfo {author} {\bibfnamefont {K.~G.}\ \bibnamefont {Derpanis}},\ and\ \bibinfo {author} {\bibfnamefont {M.~A.}\ \bibnamefont {Brubaker}},\ }\bibfield  {title} {\bibinfo {title} {Wavelet flow: Fast training of high resolution normalizing flows},\ }in\ \href {https://proceedings.neurips.cc/paper_files/paper/2020/file/4491777b1aa8b5b32c2e8666dbe1a495-Paper.pdf} {\emph {\bibinfo {booktitle} {Advances in Neural Information Processing Systems}}},\ Vol.~\bibinfo {volume} {33},\ \bibinfo {editor} {edited by\ \bibinfo {editor} {\bibfnamefont {H.}~\bibnamefont {Larochelle}}, \bibinfo {editor} {\bibfnamefont {M.}~\bibnamefont {Ranzato}}, \bibinfo {editor} {\bibfnamefont {R.}~\bibnamefont {Hadsell}}, \bibinfo {editor} {\bibfnamefont {M.}~\bibnamefont {Balcan}},\ and\ \bibinfo {editor} {\bibfnamefont {H.}~\bibnamefont {Lin}}}\ (\bibinfo  {publisher} {Curran Associates, Inc.},\ \bibinfo {year} {2020})\ pp.\ \bibinfo {pages}
  {6184--6196}\BibitemShut {NoStop}%
\bibitem [{\citenamefont {Hajimohammadi}\ \emph {et~al.}(2021)\citenamefont {Hajimohammadi}, \citenamefont {Parand},\ and\ \citenamefont {Ghodsi}}]{LDNN2021}%
  \BibitemOpen
  \bibfield  {author} {\bibinfo {author} {\bibfnamefont {Z.}~\bibnamefont {Hajimohammadi}}, \bibinfo {author} {\bibfnamefont {K.}~\bibnamefont {Parand}},\ and\ \bibinfo {author} {\bibfnamefont {A.}~\bibnamefont {Ghodsi}},\ }\href {https://arxiv.org/abs/2106.14320} {\bibinfo {title} {Legendre deep neural network (ldnn) and its application for approximation of nonlinear volterra fredholm hammerstein integral equations}} (\bibinfo {year} {2021}),\ \Eprint {https://arxiv.org/abs/2106.14320} {arXiv:2106.14320 [math.NA]} \BibitemShut {NoStop}%
\bibitem [{\citenamefont {Khan}\ \emph {et~al.}(2022)\citenamefont {Khan}, \citenamefont {Sulaiman}, \citenamefont {Kumam},\ and\ \citenamefont {Alarfaj}}]{Khan2022}%
  \BibitemOpen
  \bibfield  {author} {\bibinfo {author} {\bibfnamefont {N.~A.}\ \bibnamefont {Khan}}, \bibinfo {author} {\bibfnamefont {M.}~\bibnamefont {Sulaiman}}, \bibinfo {author} {\bibfnamefont {P.}~\bibnamefont {Kumam}},\ and\ \bibinfo {author} {\bibfnamefont {F.~K.}\ \bibnamefont {Alarfaj}},\ }\bibfield  {title} {\bibinfo {title} {Application of legendre polynomials based neural networks for the analysis of heat and mass transfer of a non-newtonian fluid in a porous channel},\ }\bibfield  {journal} {\bibinfo  {journal} {Advances in Continuous and Discrete Models}\ }\textbf {\bibinfo {volume} {2022}},\ \href {https://doi.org/10.1186/s13662-022-03676-x} {10.1186/s13662-022-03676-x} (\bibinfo {year} {2022})\BibitemShut {NoStop}%
\bibitem [{\citenamefont {Yang}\ \emph {et~al.}(2019)\citenamefont {Yang}, \citenamefont {Hou}, \citenamefont {Sun}, \citenamefont {Zhang}, \citenamefont {Weng},\ and\ \citenamefont {Luo}}]{Yang2019}%
  \BibitemOpen
  \bibfield  {author} {\bibinfo {author} {\bibfnamefont {Y.}~\bibnamefont {Yang}}, \bibinfo {author} {\bibfnamefont {M.}~\bibnamefont {Hou}}, \bibinfo {author} {\bibfnamefont {H.}~\bibnamefont {Sun}}, \bibinfo {author} {\bibfnamefont {T.}~\bibnamefont {Zhang}}, \bibinfo {author} {\bibfnamefont {F.}~\bibnamefont {Weng}},\ and\ \bibinfo {author} {\bibfnamefont {J.}~\bibnamefont {Luo}},\ }\bibfield  {title} {\bibinfo {title} {Neural network algorithm based on legendre improved extreme learning machine for solving elliptic partial differential equations},\ }\href {https://doi.org/10.1007/s00500-019-03944-1} {\bibfield  {journal} {\bibinfo  {journal} {Soft Computing}\ }\textbf {\bibinfo {volume} {24}},\ \bibinfo {pages} {1083–1096} (\bibinfo {year} {2019})}\BibitemShut {NoStop}%
\bibitem [{\citenamefont {Lee-Thorp}\ \emph {et~al.}(2022)\citenamefont {Lee-Thorp}, \citenamefont {Ainslie}, \citenamefont {Eckstein},\ and\ \citenamefont {Ontanon}}]{leethorp2022fnetmixingtokensfourier}%
  \BibitemOpen
  \bibfield  {author} {\bibinfo {author} {\bibfnamefont {J.}~\bibnamefont {Lee-Thorp}}, \bibinfo {author} {\bibfnamefont {J.}~\bibnamefont {Ainslie}}, \bibinfo {author} {\bibfnamefont {I.}~\bibnamefont {Eckstein}},\ and\ \bibinfo {author} {\bibfnamefont {S.}~\bibnamefont {Ontanon}},\ }\href {https://arxiv.org/abs/2105.03824} {\bibinfo {title} {Fnet: Mixing tokens with fourier transforms}} (\bibinfo {year} {2022}),\ \Eprint {https://arxiv.org/abs/2105.03824} {arXiv:2105.03824 [cs.CL]} \BibitemShut {NoStop}%
\bibitem [{\citenamefont {Scribano}\ \emph {et~al.}(2023)\citenamefont {Scribano}, \citenamefont {Franchini}, \citenamefont {Prato},\ and\ \citenamefont {Bertogna}}]{Scribano2023}%
  \BibitemOpen
  \bibfield  {author} {\bibinfo {author} {\bibfnamefont {C.}~\bibnamefont {Scribano}}, \bibinfo {author} {\bibfnamefont {G.}~\bibnamefont {Franchini}}, \bibinfo {author} {\bibfnamefont {M.}~\bibnamefont {Prato}},\ and\ \bibinfo {author} {\bibfnamefont {M.}~\bibnamefont {Bertogna}},\ }\bibfield  {title} {\bibinfo {title} {Dct-former: Efficient self-attention with discrete cosine transform},\ }\bibfield  {journal} {\bibinfo  {journal} {Journal of Scientific Computing}\ }\textbf {\bibinfo {volume} {94}},\ \href {https://doi.org/10.1007/s10915-023-02125-5} {10.1007/s10915-023-02125-5} (\bibinfo {year} {2023})\BibitemShut {NoStop}%
\bibitem [{\citenamefont {Budiutama}\ \emph {et~al.}(2025)\citenamefont {Budiutama}, \citenamefont {Daimon}, \citenamefont {Nishi},\ and\ \citenamefont {ichiro Matsushita}}]{budiutama2025}%
  \BibitemOpen
  \bibfield  {author} {\bibinfo {author} {\bibfnamefont {G.}~\bibnamefont {Budiutama}}, \bibinfo {author} {\bibfnamefont {S.}~\bibnamefont {Daimon}}, \bibinfo {author} {\bibfnamefont {H.}~\bibnamefont {Nishi}},\ and\ \bibinfo {author} {\bibfnamefont {Y.}~\bibnamefont {ichiro Matsushita}},\ }\href {https://arxiv.org/abs/2505.04969} {\bibinfo {title} {General transform: A unified framework for adaptive transform to enhance representations}} (\bibinfo {year} {2025}),\ \Eprint {https://arxiv.org/abs/2505.04969} {arXiv:2505.04969 [cs.LG]} \BibitemShut {NoStop}%
\bibitem [{\citenamefont {Biamonte}\ \emph {et~al.}(2017)\citenamefont {Biamonte}, \citenamefont {Wittek}, \citenamefont {Pancotti}, \citenamefont {Rebentrost}, \citenamefont {Wiebe},\ and\ \citenamefont {Lloyd}}]{biamonte2017}%
  \BibitemOpen
  \bibfield  {author} {\bibinfo {author} {\bibfnamefont {J.}~\bibnamefont {Biamonte}}, \bibinfo {author} {\bibfnamefont {P.}~\bibnamefont {Wittek}}, \bibinfo {author} {\bibfnamefont {N.}~\bibnamefont {Pancotti}}, \bibinfo {author} {\bibfnamefont {P.}~\bibnamefont {Rebentrost}}, \bibinfo {author} {\bibfnamefont {N.}~\bibnamefont {Wiebe}},\ and\ \bibinfo {author} {\bibfnamefont {S.}~\bibnamefont {Lloyd}},\ }\bibfield  {title} {\bibinfo {title} {Quantum machine learning},\ }\href {https://doi.org/10.1038/nature23474} {\bibfield  {journal} {\bibinfo  {journal} {Nature}\ }\textbf {\bibinfo {volume} {549}},\ \bibinfo {pages} {195–202} (\bibinfo {year} {2017})}\BibitemShut {NoStop}%
\bibitem [{\citenamefont {Schuld}\ and\ \citenamefont {Petruccione}(2018)}]{schuld2018}%
  \BibitemOpen
  \bibfield  {author} {\bibinfo {author} {\bibfnamefont {M.}~\bibnamefont {Schuld}}\ and\ \bibinfo {author} {\bibfnamefont {F.}~\bibnamefont {Petruccione}},\ }\href@noop {} {\emph {\bibinfo {title} {Supervised Learning with Quantum Computers}}}\ (\bibinfo  {publisher} {Springer},\ \bibinfo {address} {Cham, Switzerland},\ \bibinfo {year} {2018})\BibitemShut {NoStop}%
\bibitem [{\citenamefont {Pan}\ \emph {et~al.}(2023)\citenamefont {Pan}, \citenamefont {Lu}, \citenamefont {Wang}, \citenamefont {Hua}, \citenamefont {Xu}, \citenamefont {Li}, \citenamefont {Cai}, \citenamefont {Li}, \citenamefont {Wang}, \citenamefont {Song}, \citenamefont {Zou}, \citenamefont {Deng},\ and\ \citenamefont {Sun}}]{Pan2023}%
  \BibitemOpen
  \bibfield  {author} {\bibinfo {author} {\bibfnamefont {X.}~\bibnamefont {Pan}}, \bibinfo {author} {\bibfnamefont {Z.}~\bibnamefont {Lu}}, \bibinfo {author} {\bibfnamefont {W.}~\bibnamefont {Wang}}, \bibinfo {author} {\bibfnamefont {Z.}~\bibnamefont {Hua}}, \bibinfo {author} {\bibfnamefont {Y.}~\bibnamefont {Xu}}, \bibinfo {author} {\bibfnamefont {W.}~\bibnamefont {Li}}, \bibinfo {author} {\bibfnamefont {W.}~\bibnamefont {Cai}}, \bibinfo {author} {\bibfnamefont {X.}~\bibnamefont {Li}}, \bibinfo {author} {\bibfnamefont {H.}~\bibnamefont {Wang}}, \bibinfo {author} {\bibfnamefont {Y.-P.}\ \bibnamefont {Song}}, \bibinfo {author} {\bibfnamefont {C.-L.}\ \bibnamefont {Zou}}, \bibinfo {author} {\bibfnamefont {D.-L.}\ \bibnamefont {Deng}},\ and\ \bibinfo {author} {\bibfnamefont {L.}~\bibnamefont {Sun}},\ }\bibfield  {title} {\bibinfo {title} {Deep quantum neural networks on a superconducting processor},\ }\bibfield  {journal} {\bibinfo  {journal} {Nature Communications}\ }\textbf {\bibinfo {volume} {14}},\ \href
  {https://doi.org/10.1038/s41467-023-39785-8} {10.1038/s41467-023-39785-8} (\bibinfo {year} {2023})\BibitemShut {NoStop}%
\bibitem [{\citenamefont {Beer}\ \emph {et~al.}(2020)\citenamefont {Beer}, \citenamefont {Bondarenko}, \citenamefont {Farrelly}, \citenamefont {Osborne}, \citenamefont {Salzmann}, \citenamefont {Scheiermann},\ and\ \citenamefont {Wolf}}]{Beer2020}%
  \BibitemOpen
  \bibfield  {author} {\bibinfo {author} {\bibfnamefont {K.}~\bibnamefont {Beer}}, \bibinfo {author} {\bibfnamefont {D.}~\bibnamefont {Bondarenko}}, \bibinfo {author} {\bibfnamefont {T.}~\bibnamefont {Farrelly}}, \bibinfo {author} {\bibfnamefont {T.~J.}\ \bibnamefont {Osborne}}, \bibinfo {author} {\bibfnamefont {R.}~\bibnamefont {Salzmann}}, \bibinfo {author} {\bibfnamefont {D.}~\bibnamefont {Scheiermann}},\ and\ \bibinfo {author} {\bibfnamefont {R.}~\bibnamefont {Wolf}},\ }\bibfield  {title} {\bibinfo {title} {Training deep quantum neural networks},\ }\bibfield  {journal} {\bibinfo  {journal} {Nature Communications}\ }\textbf {\bibinfo {volume} {11}},\ \href {https://doi.org/10.1038/s41467-020-14454-2} {10.1038/s41467-020-14454-2} (\bibinfo {year} {2020})\BibitemShut {NoStop}%
\bibitem [{\citenamefont {García-Martín}\ \emph {et~al.}(2025)\citenamefont {García-Martín}, \citenamefont {Larocca},\ and\ \citenamefont {Cerezo}}]{GarcaMartn2025}%
  \BibitemOpen
  \bibfield  {author} {\bibinfo {author} {\bibfnamefont {D.}~\bibnamefont {García-Martín}}, \bibinfo {author} {\bibfnamefont {M.}~\bibnamefont {Larocca}},\ and\ \bibinfo {author} {\bibfnamefont {M.}~\bibnamefont {Cerezo}},\ }\bibfield  {title} {\bibinfo {title} {Quantum neural networks form gaussian processes},\ }\bibfield  {journal} {\bibinfo  {journal} {Nature Physics}\ }\href {https://doi.org/10.1038/s41567-025-02883-z} {10.1038/s41567-025-02883-z} (\bibinfo {year} {2025})\BibitemShut {NoStop}%
\bibitem [{\citenamefont {Preskill}(2018)}]{Preskill2018}%
  \BibitemOpen
  \bibfield  {author} {\bibinfo {author} {\bibfnamefont {J.}~\bibnamefont {Preskill}},\ }\bibfield  {title} {\bibinfo {title} {Quantum computing in the nisq era and beyond},\ }\href {https://doi.org/10.22331/q-2018-08-06-79} {\bibfield  {journal} {\bibinfo  {journal} {Quantum}\ }\textbf {\bibinfo {volume} {2}},\ \bibinfo {pages} {79} (\bibinfo {year} {2018})}\BibitemShut {NoStop}%
\bibitem [{\citenamefont {Kandala}\ \emph {et~al.}(2017)\citenamefont {Kandala}, \citenamefont {Mezzacapo}, \citenamefont {Temme}, \citenamefont {Takita}, \citenamefont {Brink}, \citenamefont {Chow},\ and\ \citenamefont {Gambetta}}]{Kandala2017}%
  \BibitemOpen
  \bibfield  {author} {\bibinfo {author} {\bibfnamefont {A.}~\bibnamefont {Kandala}}, \bibinfo {author} {\bibfnamefont {A.}~\bibnamefont {Mezzacapo}}, \bibinfo {author} {\bibfnamefont {K.}~\bibnamefont {Temme}}, \bibinfo {author} {\bibfnamefont {M.}~\bibnamefont {Takita}}, \bibinfo {author} {\bibfnamefont {M.}~\bibnamefont {Brink}}, \bibinfo {author} {\bibfnamefont {J.~M.}\ \bibnamefont {Chow}},\ and\ \bibinfo {author} {\bibfnamefont {J.~M.}\ \bibnamefont {Gambetta}},\ }\bibfield  {title} {\bibinfo {title} {Hardware-efficient variational quantum eigensolver for small molecules and quantum magnets},\ }\href {https://doi.org/10.1038/nature23879} {\bibfield  {journal} {\bibinfo  {journal} {Nature}\ }\textbf {\bibinfo {volume} {549}},\ \bibinfo {pages} {242–246} (\bibinfo {year} {2017})}\BibitemShut {NoStop}%
\bibitem [{\citenamefont {McClean}\ \emph {et~al.}(2018)\citenamefont {McClean}, \citenamefont {Boixo}, \citenamefont {Smelyanskiy}, \citenamefont {Babbush},\ and\ \citenamefont {Neven}}]{mcclean2018}%
  \BibitemOpen
  \bibfield  {author} {\bibinfo {author} {\bibfnamefont {J.~R.}\ \bibnamefont {McClean}}, \bibinfo {author} {\bibfnamefont {S.}~\bibnamefont {Boixo}}, \bibinfo {author} {\bibfnamefont {V.~N.}\ \bibnamefont {Smelyanskiy}}, \bibinfo {author} {\bibfnamefont {R.}~\bibnamefont {Babbush}},\ and\ \bibinfo {author} {\bibfnamefont {H.}~\bibnamefont {Neven}},\ }\bibfield  {title} {\bibinfo {title} {Barren plateaus in quantum neural network training landscapes},\ }\bibfield  {journal} {\bibinfo  {journal} {Nature Communications}\ }\textbf {\bibinfo {volume} {9}},\ \href {https://doi.org/10.1038/s41467-018-07090-4} {10.1038/s41467-018-07090-4} (\bibinfo {year} {2018})\BibitemShut {NoStop}%
\bibitem [{\citenamefont {Cerezo}\ \emph {et~al.}(2021)\citenamefont {Cerezo}, \citenamefont {Sone}, \citenamefont {Volkoff}, \citenamefont {Cincio},\ and\ \citenamefont {Coles}}]{cerezo2021}%
  \BibitemOpen
  \bibfield  {author} {\bibinfo {author} {\bibfnamefont {M.}~\bibnamefont {Cerezo}}, \bibinfo {author} {\bibfnamefont {A.}~\bibnamefont {Sone}}, \bibinfo {author} {\bibfnamefont {T.}~\bibnamefont {Volkoff}}, \bibinfo {author} {\bibfnamefont {L.}~\bibnamefont {Cincio}},\ and\ \bibinfo {author} {\bibfnamefont {P.~J.}\ \bibnamefont {Coles}},\ }\bibfield  {title} {\bibinfo {title} {Cost function dependent barren plateaus in shallow parametrized quantum circuits},\ }\bibfield  {journal} {\bibinfo  {journal} {Nature Communications}\ }\textbf {\bibinfo {volume} {12}},\ \href {https://doi.org/10.1038/s41467-021-21728-w} {10.1038/s41467-021-21728-w} (\bibinfo {year} {2021})\BibitemShut {NoStop}%
\bibitem [{\citenamefont {Grant}\ \emph {et~al.}(2019)\citenamefont {Grant}, \citenamefont {Wossnig}, \citenamefont {Ostaszewski},\ and\ \citenamefont {Benedetti}}]{Grant2019initialization}%
  \BibitemOpen
  \bibfield  {author} {\bibinfo {author} {\bibfnamefont {E.}~\bibnamefont {Grant}}, \bibinfo {author} {\bibfnamefont {L.}~\bibnamefont {Wossnig}}, \bibinfo {author} {\bibfnamefont {M.}~\bibnamefont {Ostaszewski}},\ and\ \bibinfo {author} {\bibfnamefont {M.}~\bibnamefont {Benedetti}},\ }\bibfield  {title} {\bibinfo {title} {An initialization strategy for addressing barren plateaus in parametrized quantum circuits},\ }\href {https://doi.org/10.22331/q-2019-12-09-214} {\bibfield  {journal} {\bibinfo  {journal} {{Quantum}}\ }\textbf {\bibinfo {volume} {3}},\ \bibinfo {pages} {214} (\bibinfo {year} {2019})}\BibitemShut {NoStop}%
\bibitem [{\citenamefont {Sugawara}\ \emph {et~al.}(2025)\citenamefont {Sugawara}, \citenamefont {Inomata}, \citenamefont {Okubo},\ and\ \citenamefont {Todo}}]{sugawara2025embeddingtreetensornetworks}%
  \BibitemOpen
  \bibfield  {author} {\bibinfo {author} {\bibfnamefont {S.}~\bibnamefont {Sugawara}}, \bibinfo {author} {\bibfnamefont {K.}~\bibnamefont {Inomata}}, \bibinfo {author} {\bibfnamefont {T.}~\bibnamefont {Okubo}},\ and\ \bibinfo {author} {\bibfnamefont {S.}~\bibnamefont {Todo}},\ }\href {https://arxiv.org/abs/2501.18856} {\bibinfo {title} {Embedding of tree tensor networks into shallow quantum circuits}} (\bibinfo {year} {2025}),\ \Eprint {https://arxiv.org/abs/2501.18856} {arXiv:2501.18856 [quant-ph]} \BibitemShut {NoStop}%
\bibitem [{\citenamefont {Skolik}\ \emph {et~al.}(2021)\citenamefont {Skolik}, \citenamefont {McClean}, \citenamefont {Mohseni}, \citenamefont {van~der Smagt},\ and\ \citenamefont {Leib}}]{Skolik2021}%
  \BibitemOpen
  \bibfield  {author} {\bibinfo {author} {\bibfnamefont {A.}~\bibnamefont {Skolik}}, \bibinfo {author} {\bibfnamefont {J.~R.}\ \bibnamefont {McClean}}, \bibinfo {author} {\bibfnamefont {M.}~\bibnamefont {Mohseni}}, \bibinfo {author} {\bibfnamefont {P.}~\bibnamefont {van~der Smagt}},\ and\ \bibinfo {author} {\bibfnamefont {M.}~\bibnamefont {Leib}},\ }\bibfield  {title} {\bibinfo {title} {Layerwise learning for quantum neural networks},\ }\bibfield  {journal} {\bibinfo  {journal} {Quantum Machine Intelligence}\ }\textbf {\bibinfo {volume} {3}},\ \href {https://doi.org/10.1007/s42484-020-00036-4} {10.1007/s42484-020-00036-4} (\bibinfo {year} {2021})\BibitemShut {NoStop}%
\bibitem [{\citenamefont {Budiutama}\ \emph {et~al.}(2024)\citenamefont {Budiutama}, \citenamefont {Daimon}, \citenamefont {Nishi}, \citenamefont {Kaneko}, \citenamefont {Ohtsuki},\ and\ \citenamefont {Matsushita}}]{budiutama2024}%
  \BibitemOpen
  \bibfield  {author} {\bibinfo {author} {\bibfnamefont {G.}~\bibnamefont {Budiutama}}, \bibinfo {author} {\bibfnamefont {S.}~\bibnamefont {Daimon}}, \bibinfo {author} {\bibfnamefont {H.}~\bibnamefont {Nishi}}, \bibinfo {author} {\bibfnamefont {R.}~\bibnamefont {Kaneko}}, \bibinfo {author} {\bibfnamefont {T.}~\bibnamefont {Ohtsuki}},\ and\ \bibinfo {author} {\bibfnamefont {Y.-i.}\ \bibnamefont {Matsushita}},\ }\bibfield  {title} {\bibinfo {title} {Channel attention for quantum convolutional neural networks},\ }\href {https://doi.org/10.1103/PhysRevA.110.012447} {\bibfield  {journal} {\bibinfo  {journal} {Phys. Rev. A}\ }\textbf {\bibinfo {volume} {110}},\ \bibinfo {pages} {012447} (\bibinfo {year} {2024})}\BibitemShut {NoStop}%
\bibitem [{\citenamefont {Smith}\ \emph {et~al.}(2022)\citenamefont {Smith}, \citenamefont {Jobst}, \citenamefont {Green},\ and\ \citenamefont {Pollmann}}]{smith_crossing_2022}%
  \BibitemOpen
  \bibfield  {author} {\bibinfo {author} {\bibfnamefont {A.}~\bibnamefont {Smith}}, \bibinfo {author} {\bibfnamefont {B.}~\bibnamefont {Jobst}}, \bibinfo {author} {\bibfnamefont {A.~G.}\ \bibnamefont {Green}},\ and\ \bibinfo {author} {\bibfnamefont {F.}~\bibnamefont {Pollmann}},\ }\bibfield  {title} {\bibinfo {title} {Crossing a topological phase transition with a quantum computer},\ }\href {https://doi.org/10.1103/PhysRevResearch.4.L022020} {\bibfield  {journal} {\bibinfo  {journal} {Physical Review Research}\ }\textbf {\bibinfo {volume} {4}},\ \bibinfo {pages} {L022020} (\bibinfo {year} {2022})}\BibitemShut {NoStop}%
\bibitem [{\citenamefont {Bertlmann}\ and\ \citenamefont {Krammer}(2008)}]{bertlmann_bloch_2008}%
  \BibitemOpen
  \bibfield  {author} {\bibinfo {author} {\bibfnamefont {R.~A.}\ \bibnamefont {Bertlmann}}\ and\ \bibinfo {author} {\bibfnamefont {P.}~\bibnamefont {Krammer}},\ }\bibfield  {title} {\bibinfo {title} {Bloch vectors for qudits},\ }\href {https://doi.org/10.1088/1751-8113/41/23/235303} {\bibfield  {journal} {\bibinfo  {journal} {Journal of Physics A: Mathematical and Theoretical}\ }\textbf {\bibinfo {volume} {41}},\ \bibinfo {pages} {235303} (\bibinfo {year} {2008})}\BibitemShut {NoStop}%
\bibitem [{\citenamefont {Verresen}\ \emph {et~al.}(2017)\citenamefont {Verresen}, \citenamefont {Moessner},\ and\ \citenamefont {Pollmann}}]{PhysRevB.96.165124}%
  \BibitemOpen
  \bibfield  {author} {\bibinfo {author} {\bibfnamefont {R.}~\bibnamefont {Verresen}}, \bibinfo {author} {\bibfnamefont {R.}~\bibnamefont {Moessner}},\ and\ \bibinfo {author} {\bibfnamefont {F.}~\bibnamefont {Pollmann}},\ }\bibfield  {title} {\bibinfo {title} {One-dimensional symmetry protected topological phases and their transitions},\ }\href {https://doi.org/10.1103/PhysRevB.96.165124} {\bibfield  {journal} {\bibinfo  {journal} {Phys. Rev. B}\ }\textbf {\bibinfo {volume} {96}},\ \bibinfo {pages} {165124} (\bibinfo {year} {2017})}\BibitemShut {NoStop}%
\bibitem [{\citenamefont {Verresen}\ \emph {et~al.}(2018)\citenamefont {Verresen}, \citenamefont {Jones},\ and\ \citenamefont {Pollmann}}]{PhysRevLett.120.057001}%
  \BibitemOpen
  \bibfield  {author} {\bibinfo {author} {\bibfnamefont {R.}~\bibnamefont {Verresen}}, \bibinfo {author} {\bibfnamefont {N.~G.}\ \bibnamefont {Jones}},\ and\ \bibinfo {author} {\bibfnamefont {F.}~\bibnamefont {Pollmann}},\ }\bibfield  {title} {\bibinfo {title} {Topology and edge modes in quantum critical chains},\ }\href {https://doi.org/10.1103/PhysRevLett.120.057001} {\bibfield  {journal} {\bibinfo  {journal} {Phys. Rev. Lett.}\ }\textbf {\bibinfo {volume} {120}},\ \bibinfo {pages} {057001} (\bibinfo {year} {2018})}\BibitemShut {NoStop}%
\bibitem [{\citenamefont {Leone}\ \emph {et~al.}(2024)\citenamefont {Leone}, \citenamefont {Oliviero}, \citenamefont {Cincio},\ and\ \citenamefont {Cerezo}}]{Leone2024practicalusefulness}%
  \BibitemOpen
  \bibfield  {author} {\bibinfo {author} {\bibfnamefont {L.}~\bibnamefont {Leone}}, \bibinfo {author} {\bibfnamefont {S.~F.}\ \bibnamefont {Oliviero}}, \bibinfo {author} {\bibfnamefont {L.}~\bibnamefont {Cincio}},\ and\ \bibinfo {author} {\bibfnamefont {M.}~\bibnamefont {Cerezo}},\ }\bibfield  {title} {\bibinfo {title} {On the practical usefulness of the {H}ardware {E}fficient {A}nsatz},\ }\href {https://doi.org/10.22331/q-2024-07-03-1395} {\bibfield  {journal} {\bibinfo  {journal} {{Quantum}}\ }\textbf {\bibinfo {volume} {8}},\ \bibinfo {pages} {1395} (\bibinfo {year} {2024})}\BibitemShut {NoStop}%
\end{thebibliography}%


\end{document}